\pdfoutput=1
\documentclass[journal,10pt]{IEEEtran}

\usepackage{amsmath,amssymb}
\usepackage{array,booktabs,tabularx}
\usepackage{cite}
\usepackage[T1]{fontenc}
\usepackage{microtype}
\usepackage{newtxtext,newtxmath}
\usepackage{rotating}
\usepackage{url}
\usepackage[hidelinks]{hyperref}
\usepackage{placeins}

\newcolumntype{Y}{>{\raggedright\arraybackslash}X}
\title{Predictive Exposure and Cryptographic Readiness: A Vendor-Neutral Framework, Bounded Multivocal Evidence Review, and Reproducible Synthetic Evaluation for SD-WAN Environments}

\author{Saeed~Alam \\ \textit{School of Computing Sciences and Computer Engineering, University of Southern Mississippi, Hattiesburg, MS, USA}}

\begin{document}

\maketitle

\begin{abstract}
SD-WAN teams often use static severity scores to decide what to fix first. These scores do not show live exploitation, network exposure, attack paths, business impact, or cryptographic migration risk. This study asks whether a vendor-neutral PECR framework can produce a different and more traceable ranking than CVSS alone. We reviewed 19 DOI-verified publications from 2020--2026 and five current NIST documents. The review supported ten normalized factors, one weighted score, and a separate confidence measure. We then compared equal-weight PECR with CVSS in five synthetic cases. The tests used Kendall's \(\tau_b\), mean absolute rank shift (\(\mathrm{MARS}\)), top-three Jaccard overlap (\(J_3\)), factor removal, and bounded weight changes. CVSS ranked the cases A--D--B--E--C. PECR ranked them A--B--E--C--D. The results were \(\tau_b=0.40\), \(\mathrm{MARS}=1.2\), and \(J_3=0.50\). The exact \(\tau\) test was not significant (\(p=0.483\)) because the sample had only five cases. The PECR order remained unchanged in 87.5\% of 1,024 weight combinations. It also remained unchanged in eight of ten single-factor removal tests. The evidence supports separate measures for severity, exploitation, and organizational context. Evidence for live SD-WAN attack paths and daily PQC triage is still limited. PECR can produce a different and auditable ranking. This synthetic test does not prove better operational results or better human understanding.
\end{abstract}


\section{Introduction}\label{sec:introduction}

Vulnerability teams must decide what to fix first while staff time and maintenance windows are limited. CVSS describes technical severity. It does not measure exploitation probability or the effect on a specific asset \cite{ref4}, \cite{ref12}, \cite{ref13}. In SD-WAN, priority may also depend on Internet exposure, control-plane access, segmentation, business function, and cryptographic dependencies. Past research has not shown how to combine all these signals in one vendor-neutral ranking.

PECR combines ten types of information. They cover severity, exploitation, exposure, attack paths, operational impact, and cryptographic readiness. This paper does not claim that ten is the perfect number. Equal weights are used only as a clear baseline for the synthetic test. Real deployments must calibrate weights against measured outcomes. The aim is to define the factors, data sources, rules, and tests without making claims beyond the evidence.

This question draws on two types of evidence. Research papers provide methods and measured results. Standards bodies provide scoring rules, cryptographic requirements, and migration guidance. PRISMA 2020 guided the transparency checks \cite{ref1}. However, the available records do not support a PRISMA flow diagram. Software-engineering guidance helped define eligibility \cite{ref2}. Academic and grey-literature sources were assessed separately \cite{ref3}. Vendor advisories may explain a product issue, but they do not prove general effectiveness.

\section{Refined Research Questions}\label{sec:research-questions}

The evidence review and framework evaluation address different questions. The review identifies and bounds support for candidate factors. The synthetic evaluation tests calculation mechanics and rank sensitivity. Operational validation still requires comparison on matched vulnerability--asset instances with an external outcome.

\subsection{Primary review question}

PRQ. Which evidence-supported factors, data sources, and measurement methods can be combined into a vendor-neutral model for prioritizing vulnerability remediation in SD-WAN environments?

\subsection{Supporting review questions}

RQ1. Which static and dynamic factors are used by existing vulnerability prioritization approaches, and how are those factors measured?

RQ2. What evidence supports the use of observed exploitation, predicted exploitation, live exposure, and telemetry freshness in remediation decisions?

RQ3. How do existing approaches represent attainable privilege, attack-path reachability, lateral-movement potential, blast radius, operational consequence, and systemic concentration?

RQ4. Which characteristics of SD-WAN and SASE environments change the meaning or measurement of vulnerability priority?

RQ5. How are cryptographic dependencies, crypto-agility, confidentiality lifetime, post-quantum migration urgency, and harvest-now-decrypt-later exposure represented in security decisions?

RQ6. Which normalization, weighting, uncertainty, missing-data, confidence, and sensitivity-analysis methods are suitable for a transparent multi-factor prioritization model?

\subsection{Framework evaluation question}

ERQ. When PECR uses the ten selected factors, does it rank the same cases differently from CVSS alone? Which factors cause the changes? Do the rankings remain stable when weights, data age, or missing values change?

The ERQ uses the word ``traceable'' instead of ``more explainable.'' A traceable score shows its inputs, sources, and factor contributions. This does not prove that analysts understand or trust the score. A separate analyst study is needed before making that claim.

\section{Contribution and Claim Boundary}\label{sec:claim-boundary}

The study separates evidence synthesis, framework design, scenario evaluation, and external validation. Table~\ref{tab:claim-boundary} defines what each phase can support. This separation prevents a literature map or a synthetic demonstration from being presented as proof of operational effectiveness.

\begin{table*}[!t]
\centering
\caption{Contribution and Claim Boundary}
\label{tab:claim-boundary}
\scriptsize
\setlength{\tabcolsep}{3pt}
\renewcommand{\arraystretch}{1.12}
\begin{tabularx}{\textwidth}{@{}p{0.12\textwidth}YYY@{}}
\toprule
\textbf{Phase} & \textbf{Produces} & \textbf{Supports} & \textbf{Does not establish} \\
\midrule
1. Bounded review & Evidence map; factor taxonomy; measurement and evaluation methods; gaps & A traceable rationale for selecting or rejecting candidate PECR factors & That PECR improves real remediation outcomes \\
2. Framework design & Ten-factor model; definitions; normalization; confidence and missing-data rules & A reproducible and vendor-neutral research artifact & Empirical validity or optimal weights \\
3. Scenario evaluation & CVSS and PECR rankings; rank metrics; ablation and sensitivity results & Feasibility, internal consistency, rank difference, and decision traceability & Field effectiveness or generalizability when the data are synthetic \\
4. External validation & Operational datasets, analyst assessments, and longitudinal outcomes & Claims about usefulness, calibration, and human explainability & Part of the present review unless such data are collected \\
\bottomrule
\end{tabularx}
\end{table*}

\section{Review and Evaluation Method}\label{sec:review-method}

PICOC was used as a coding boundary for the supplied and DOI-expanded evidence corpus (Table~\ref{tab:picoc}). It kept transferable methods visible while preventing enterprise-security studies from being treated as direct SD-WAN evidence.

\begin{table*}[!t]
\centering
\caption{PICOC Definition for the PECR Review}
\label{tab:picoc}
\scriptsize
\setlength{\tabcolsep}{3pt}
\renewcommand{\arraystretch}{1.12}
\begin{tabularx}{\textwidth}{@{}p{0.13\textwidth}YY@{}}
\toprule
\textbf{Element} & \textbf{Definition} & \textbf{Operational scope} \\
\midrule
Population & Vulnerabilities and affected assets in SD-WAN, SASE, SDN, and closely transferable enterprise-network environments & Controllers, orchestrators, management services, branch edges, gateways, identity services, segmentation controls, and cryptographic dependencies. General endpoint-only studies are included only when their method is transferable. \\
Intervention & Context-aware vulnerability prioritization & Methods may use severity, exploitation evidence, probability, exposure, privilege, reachability, attack paths, blast radius, consequence, concentration, or cryptographic readiness. \\
Comparison & Severity-only and other baseline ranking methods & CVSS-only ranking is the primary baseline. EPSS-, KEV-, SSVC-, asset-criticality-, and attack-graph-based approaches are secondary comparators when reported. \\
Outcomes & Evidence needed to design and assess a prioritization model & Factor definitions; data sources; normalization; weights; uncertainty; missing-data rules; rank changes; decision traceability; sensitivity; reproducibility; and operational limitations. \\
Context & Vendor-neutral enterprise remediation decisions & Distributed branch networks with centralized policy or control, direct Internet exposure, segmentation, and cryptographic trust dependencies. Product-specific evidence may inform a mechanism but cannot define the framework. \\
\bottomrule
\end{tabularx}
\end{table*}

\subsection{Eligibility, time, and source boundaries}

The bounded corpus contains English-language topical publications dated 2020--2026 that were supplied in the source CSV or added through DOI-led citation expansion. Methodological guidance and current standards were retained separately from the 19 topical publications so they did not inflate the empirical evidence count. Journal and conference publications, a preprint, a technical paper, and authoritative NIST publications were coded as separate source types. Only English-language records were reviewed; this restriction limits geographic and linguistic coverage. Standards and guidance were checked for publication status and version on 29 July 2026; time-sensitive sources require another check immediately before submission.

\subsection{Evidence identification and selection}

The evidence set began with eight DOI records from the supplied CSV. DOI checks and citation searches added eleven non-duplicate studies. This produced 19 topical publications. We checked each DOI and used the publisher page or an author manuscript when available. We also added five NIST documents for cryptographic readiness. They were FIPS 203, FIPS 204, FIPS 205, NIST IR 8547 Initial Public Draft, and NIST CSWP 39upd1 \cite{ref23,ref24,ref25,ref26,ref27}. The source files did not include a database search log or initial retrieval totals. They also did not include duplicate counts, full-text exclusions, or a second-reviewer record. We did not reconstruct these numbers. For this reason, the paper reports a bounded multivocal review rather than a complete PRISMA review. This wording makes clear that the 19-study set is not exhaustive.

\subsection{Structured appraisal and extraction}

One reviewer used a six-item quality check. It covered source status, method clarity, evaluation evidence, unit relevance, reproducibility, and claim support. Each item received a score from 0 to 2. Total scores of 9--12 were high confidence. Scores of 6--8 were moderate, and 0--5 were limited. We extracted the source type, unit, setting, sample size, data date, factors, comparator, outcomes, uncertainty, and limitations. A second reviewer did not repeat this assessment. The ratings are therefore used to limit claims, not to estimate pooled effects.

\subsection{Ethics and reproducibility}

The review used published sources, and the test used synthetic values. We did not collect personal data, production telemetry, or confidential company data. Human-subject review was therefore not needed for this study. A future analyst or operational study may need institutional approval. It would also need consent, de-identification, access controls, and a data-management plan.

\subsection{Synthetic framework evaluation}

PECR and CVSS ranked the same five synthetic vulnerability--asset instances. Rank divergence was measured with Kendall's \(\tau_b\), mean absolute rank shift (\(\mathrm{MARS}\)), and top-k Jaccard overlap (\(J_k\)). Equations~\eqref{eq:mars} and~\eqref{eq:jaccard} define the latter two measures, where r is rank position and T is the set of top-k instances.

\begin{equation}
\mathrm{MARS} = \frac{1}{n}\sum_{i=1}^{n}\left|r_i^{\mathrm{PECR}}-r_i^{\mathrm{CVSS}}\right|
\label{eq:mars}
\end{equation}

\begin{equation}
J_k = \frac{\left|T_k^{\mathrm{PECR}}\cap T_k^{\mathrm{CVSS}}\right|}{\left|T_k^{\mathrm{PECR}}\cup T_k^{\mathrm{CVSS}}\right|}
\label{eq:jaccard}
\end{equation}

The test compared every pair of cases and removed one factor at a time. It also tested all 1,024 weight combinations made by changing each weight by \(\pm 20\%\) and then normalizing the weights. A different ranking does not mean a better ranking. The synthetic cases have no external outcome. They cannot prove avoided loss, better exploit prediction, useful remediation, or faster risk reduction. We assessed traceability only through complete records and visible score components.

\section{Literature Synthesis from the DOI-Verified Corpus}\label{sec:literature-synthesis}

The supplied CSV contained eight studies published from 2022 to 2026. DOI checks and citation searches added eleven studies from 2020 to 2026. We checked each reference through its DOI. We also checked important claims against a publisher page or author manuscript when available. The final set contains 19 studies, but it is not exhaustive. It supports framework design, not a complete PRISMA review.

\subsection{Severity and exploitation are related but distinct signals}

Tan et al. found no clear correlation between CVSS severity and EPSS probability \cite{ref4}. The two measures represent different constructs: potential technical impact and predicted exploitation. PECR should preserve that distinction and treat exploit probability as a dated, time-varying input.

Jacobs et al. presented EPSS as an open, data-based estimate of exploit probability \cite{ref12}. Their 2021 model predicted exploitation during the first 12 months after disclosure and reported \(\mathrm{ROC\ AUC}=0.838\). Later EPSS versions use a 30-day period \cite{ref13}. A PECR record should store the model version, date, score, and percentile. A score without this time information can be misleading.

Koscinski et al. compared CVSS, EPSS, SSVC, and the Exploitability Index for 600 Microsoft Patch Tuesday vulnerabilities \cite{ref13}. Agreement between the systems was weak. Only five vulnerabilities appeared in every top-100 list. Large groups of tied scores also reduced ranking value. The study shows that accepted systems can produce very different orders. It does not show that combining them always gives a better result. PECR must test its combined ranking against an external outcome.

Allodi et al. studied CVSS v3 ratings from 73 people across 30 vulnerabilities \cite{ref14}. People with security knowledge were 30\%--60\% less likely to make an error than untrained participants. Experienced professionals did not show a clear advantage over security-focused students. This result shows that CVSS includes important human judgment. PECR should show the source, confidence, and missing-data status for each assessed value.

Iannone et al. tested 72 early exploit-prediction settings and five pretrained language models \cite{ref15}. They used early CVE descriptions and linked discussions. In the time-aware test, CVE descriptions and SecurityFocus discussions gave the most useful signals. Pretrained language models performed poorly without added security training. PECR should treat early predictions as uncertain evidence, not as proof of exploitation.

Shimizu and Hashimoto tested a two-stage process \cite{ref5}. It first used KEV membership or an EPSS threshold, and then applied a CVSS threshold. Their test covered 28,377 vulnerabilities and reported \(\mathrm{AUC}=0.8678\). At an EPSS threshold of 0.088, the true-positive rate was 48.89\%. The reported workload reduction was 96.9\%, but the classes were highly imbalanced. This result shows an important trade-off. A smaller urgent queue may still miss exploited vulnerabilities.

\subsection{Organizational context changes remediation order}

Ahmadi Mehri, Arlos, and Casalicchio built a context-aware method \cite{ref6}. Organizations choose the criteria and weights based on their risk needs. The test produced different remediation orders when those choices changed. Local context can therefore change priority. However, local weights must be documented, reviewed, and tested for sensitivity.

Dubey and Maity combined semantic representations with asset importance, exposure, and exploitation likelihood \cite{ref7}. They assessed the remediation order with Mean Reciprocal Rank and Recall@K instead of relying only on classification accuracy. Those measures match PECR's intended output: an ordered queue. Source attribution and decomposed score components also improve auditability, although they do not establish human comprehension.

Agyei et al. combined CVSS, EPSS, KEV, asset criticality, Internet exposure, and compensating controls in a weighted hybrid-cloud model \cite{ref8}. Because the evaluation used simulated data, the reported reductions in urgent work are scenario results rather than evidence of field effectiveness. The study's relevant design features are factor-level contributions, rule-based overrides, and audit-readable decision records.

\subsection{Threat context is useful only when mappings are complete}

Shreyas and Arun Kumar integrated CVSS, EPSS, and MITRE ATT\&CK in static and adaptive models \cite{ref9}. EPSS carried the strongest learned weight, whereas ATT\&CK contributed little or received a negative weight when mappings were sparse. A conceptually attractive factor should not receive a large weight when coverage is poor. PECR should measure coverage, freshness, and mapping confidence separately from the factor value.

Anwer and Ali analyzed 1,577 KEV records across 257 vendors using exploitation volume, ransomware association, and remediation delay \cite{ref10}. They reported Spearman correlations above 0.96 across alternative weight settings. The analysis supports observed-exploitation signals and formal sensitivity testing, but its principal unit is the vendor. PECR must rank a vulnerability on a specific asset and attack path.

Sri's practitioner framework combines CVSS, EPSS, KEV, and IT asset-management context for small and medium-sized enterprises \cite{ref11}. Its implementation guidance is relevant, but the stated workload reduction is an estimate rather than a reproduced experiment. The source should remain in the grey-literature stratum and should not carry the same evidentiary weight as a peer-reviewed evaluation.

Briliyant et al. applied live EPSS values and an SSVC decision tree to 14 IoT vulnerability cases \cite{ref16}. They identified three cases requiring immediate attention that a static method missed. The sample is small and outside SD-WAN, and the abstract reports internally inconsistent percentages. This review therefore retains the count but not the disputed percentages.

Chhillar et al. proposed a two-stage risk-scoring engine \cite{ref17}. It combines CVSS, EPSS, KEV, threat intelligence, and internal telemetry. The first stage predicts exploitability. The second stage creates a contextual score. The design is relevant to PECR. However, the six-page paper does not provide enough evaluation detail to confirm its claimed alert reduction.

Moraghebi and Ali analyzed more than 1,600 CISA KEV records and separated ransomware-linked vulnerabilities from the broader exploited set \cite{ref18}. The ransomware-linked records showed concentrated vendor exposure, greater operational urgency, and changing remediation windows. These findings motivate a separate consequence signal, but the study examines catalog, vendor, and temporal patterns rather than asset-specific SD-WAN impact.

Yu et al. studied vulnerability co-exploitation with a large knowledge graph \cite{ref19}. The graph had more than 355,000 nodes and four million links. Their model predicts vulnerabilities that may be exploited together. This supports a separate dependency factor. It does not measure live reachability, attainable privilege, or lateral movement in an SD-WAN network.

\subsection{AI integration, process context, and cross-domain transferability}

Siewruk and Berej reviewed generative and learning models across code, binary, dependency, graph, and deployment representations \cite{ref20}. Their risk-aware model combines likelihood, exposure, and impact while emphasizing guardrails, retained evidence, triage time, developer acceptance, and change-failure rate. These are useful adoption measures. As a review and design overview, however, the article is not independent evidence that a particular scoring model improves outcomes.

Lysetskyi frames vulnerability management as a continuous sequence of detection, normalization, prioritization, remediation, verification, and reporting \cite{ref21}. The paper argues that scanners and CVE records are necessary but insufficient without environmental context, predictive models, and threat analytics. It provides process background, not comparative evidence for a weighting formula.

Oka and Vadamalu examine CVSS, EPSS, and SSVC in software-defined vehicles, where external connectivity and software complexity expand the attack surface \cite{ref22}. Their lifecycle perspective is transferable to software-defined network edges. The automotive setting cannot establish SD-WAN factor values, but it reinforces the distinction among severity, predicted exploitation, and mission context.

\subsection{Evidence map and remaining gap}

\begin{sidewaystable*}[!t]
\centering
\caption{DOI-Verified Corpus and Its Bounded Contribution}
\label{tab:evidence-corpus}
\tiny
\setlength{\tabcolsep}{3pt}
\renewcommand{\arraystretch}{1.12}
\begin{tabularx}{\textheight}{@{}p{0.12\textheight}YYY@{}}
\toprule
\textbf{Study} & \textbf{Evidence basis} & \textbf{Primary signals} & \textbf{PECR contribution and limit} \\
\midrule
\cite{ref4} Tan et al., 2025 & Peer-reviewed IEEE conference analysis & CVSS, EPSS, score movement & Supports separating severity from exploitation likelihood; no asset, topology, or operational context. \\
\cite{ref5} Shimizu \& Hashimoto, 2026 & Peer-reviewed IEEE Access evaluation; 28,377 vulnerabilities & KEV, EPSS, CVSS thresholds & Strong multi-signal triage evidence; threshold trade-offs remain; no live network telemetry or cryptographic factor. \\
\cite{ref6} Ahmadi Mehri et al., 2022 & Journal prototype and scenario comparison & Organizational weights, risk appetite, vulnerability age & Supports context-specific ordering; locally selected weights require governance and sensitivity testing. \\
\cite{ref7} Dubey \& Maity, 2026 & Peer-reviewed IEEE conference learning-to-rank study & Asset importance, exposure, exploit likelihood, semantic context & Supports ranking metrics and decomposed evidence; general enterprise scope, not SD-WAN-specific. \\
\cite{ref8} Agyei et al., 2026 & Journal article using simulated hybrid-cloud data & CVSS, EPSS, KEV, criticality, exposure, controls & Useful transparent design pattern; simulated outcomes do not establish field effectiveness or human explainability. \\
\cite{ref9} Shreyas \& Arun Kumar, 2025 & Peer-reviewed IEEE conference adaptive model & CVSS, EPSS, ATT\&CK mappings & Shows the effect of sparse threat mappings; factor value depends on coverage and data quality. \\
\cite{ref10} Anwer \& Ali, 2026 & Journal analytics study; 1,577 KEV records and 257 vendors & KEV, ransomware association, remediation delay & Supports operational exploitation signals and sensitivity analysis; vendor-level output differs from PECR's vulnerability-asset unit. \\
\cite{ref11} Sri, 2026 & SSRN practitioner preprint & CVSS, EPSS, KEV, ITAM asset context & Provides adoption guidance; estimated results and no peer-review status limit evidentiary weight. \\
\cite{ref12} Jacobs et al., 2021 & Peer-reviewed foundational EPSS article & Exploit probability; historical 12-month horizon & Supports a separate predictive signal; model version and horizon must be recorded. \\
\cite{ref13} Koscinski et al., 2025 & ACM CCS study; 600 Patch Tuesday CVEs & CVSS, EPSS, SSVC, Exploitability Index & Direct rank-divergence evidence; weak agreement warns against unvalidated score fusion. \\
\cite{ref14} Allodi et al., 2020 & Journal experiment; 73 assessors, 30 CVEs & CVSS accuracy, expertise, confidence & Shows assessor-dependent variance; supports provenance and confidence fields. \\
\cite{ref15} Iannone et al., 2024 & Journal time-aware prediction study & Initial CVE text, discussions, ML, LLMs & Supports early prediction with uncertainty; pretrained LLM performance was limited. \\
\cite{ref16} Briliyant et al., 2025 & IEEE conference; 14 IoT cases & Live EPSS, SSVC, exploitation status & Shows cross-domain rank changes; small sample and inconsistent abstract percentages limit inference. \\
\cite{ref17} Chhillar et al., 2025 & Six-page IEEE conference framework & CVSS, EPSS, KEV, threat intelligence, telemetry & Relevant telemetry architecture; accessible evidence is insufficient for effectiveness claims. \\
\cite{ref18} Moraghebi \& Ali, 2026 & Journal analysis; more than 1,600 KEV records & Ransomware, vendor exposure, remediation windows & Supports operational consequence; catalog/vendor analysis is not asset-level validation. \\
\cite{ref19} Yu et al., 2025 & IEEE knowledge-graph experiment & Co-exploitation, semantic and structural links & Supports multi-CVE dependency analysis; does not measure live network attack paths. \\
\cite{ref20} Siewruk \& Berej, 2026 & Peer-reviewed IEEE Access review & AI models, likelihood, exposure, impact, guardrails & Useful adoption and audit measures; secondary evidence, not independent validation. \\
\cite{ref21} Lysetskyi, 2026 & Journal process overview & Context, predictive analytics, continuous VM cycle & Provides process background; no comparative ranking evaluation. \\
\cite{ref22} Oka \& Vadamalu, 2026 & SAE technical paper; automotive attack analysis & CVSS, EPSS, SSVC, mission context & Transferable software-defined edge analogy; not direct SD-WAN evidence. \\
\bottomrule
\end{tabularx}
\end{sidewaystable*}
\FloatBarrier

Six design needs appeared across the 19 studies. Severity and exploit likelihood should remain separate. Predictive scores need a version and date. Confirmed exploitation should trigger urgent review. Asset, mission, and operational context should affect priority. The model should also show multi-vulnerability links, data sources, and confidence. Existing scoring systems can produce very different rankings. However, the evidence does not validate SD-WAN attack paths, lateral movement, control-plane concentration, or telemetry-age rules. It also does not validate a C-BOM or daily post-quantum migration score. Finally, the studies do not prove that ten factors or one weight set is best.

\section{Bounded Review Results}\label{sec:review-results}

Evidence limit. The available sources support a structured review with checked DOIs. They do not support reconstructed PRISMA counts, exhaustive coverage, or reviewer-agreement statistics. The results in this section apply only to the 19 studies and five NIST documents described in Section~\ref{sec:review-method}.

\subsection{Evidence identification and corpus boundary}

The evidence set was assembled through supplied records, DOI resolution, and citation-led expansion. Counts that were not present in the source files are reported as unavailable rather than inferred.

\noindent\textbf{Registration:} None. The review was completed retrospectively as a bounded evidence synthesis and is not presented as a preregistered SLR.

\noindent\textbf{Evidence-status date:} DOI identities and current NIST publication status were checked through 29 July 2026.

\noindent\textbf{Identification:} Eight supplied DOI records plus eleven DOI-led additions produced 19 topical publications; five NIST publications were added as a separate standards stratum.

\noindent\textbf{Duplicate and exclusion counts:} Unavailable because no database export or screening log was supplied. No count is reconstructed.

\noindent\textbf{Included corpus:} 19 topical publications were retained for design synthesis; 27 total references include three review-method sources and five NIST publications.

\noindent\textbf{Full-text exclusions:} Not reportable from the available records. Abstract-only limitations are identified in the evidence synthesis where applicable.

\noindent\textbf{Reviewer agreement:} Not applicable. One reviewer performed the evidence audit; no independent screening or adjudication record exists.

\noindent\textbf{PRISMA diagram:} Intentionally omitted. A flow diagram would imply retrieval and screening data that were not collected.

\subsection{Included-source characteristics}

We summarized the studies by year, source type, setting, design, unit of analysis, comparator, and outcome. Primary studies, reviews, and official guidance were kept separate. This avoids counting the same evidence twice.

\noindent\textbf{Evidence composition:} 19 topical publications: 17 journal or conference papers, one SSRN preprint \cite{ref11}, and one SAE technical paper \cite{ref22}. Five NIST standards or guidance publications form a separate authoritative stratum \cite{ref23,ref24,ref25,ref26,ref27}.

\noindent\textbf{Directness:} No included study validates vulnerability prioritization on an operational SD-WAN or SASE dataset. All 19 topical publications are transferred from general vulnerability management, enterprise or hybrid cloud, IoT, automotive, software engineering, or threat-knowledge settings.

\noindent\textbf{Study design:} 14 records report an empirical, prototype, case, analytics, or simulated evaluation; five are review, framework, process, practitioner, or technical treatments. The five NIST records define standards or transition guidance.

\noindent\textbf{Coverage:} Topical publications span 2020--2026. Sectors include enterprise IT, hybrid cloud, IoT, automotive systems, and cross-domain vulnerability management; geographic representativeness cannot be inferred from this corpus.

\subsection{Quality appraisal and confidence in the evidence}

The quality check limited how each source could support a factor, threshold, weight, or effectiveness claim. We report confidence by source type. We also checked whether the main conclusions changed after removing limited-confidence sources.

\noindent\textbf{Appraisal method:} A retrospective six-item, 0--12 rubric assessed source status, methodological transparency, evaluation evidence, unit relevance, reproducibility, and claim support. The rubric was not preregistered or independently piloted.

\noindent\textbf{Confidence distribution:} Seven studies were high confidence \cite{ref5}, \cite{ref6}, \cite{ref12,ref13,ref14,ref15}, \cite{ref19}. Eight were moderate \cite{ref4}, \cite{ref7}, \cite{ref9}, \cite{ref10}, \cite{ref16}, \cite{ref18}, \cite{ref20}, \cite{ref22}. Four were limited \cite{ref8}, \cite{ref11}, \cite{ref17}, \cite{ref21}. These ratings only control how the sources are used in this paper. They are not rankings of journal quality.

\noindent\textbf{Use after appraisal:} No record was removed from qualitative design synthesis. Limited-confidence records were used only for architecture, process, or implementation context and not to justify effect sizes, thresholds, or weights.

\noindent\textbf{Sensitivity:} Removing the four limited-confidence studies did not change the main findings. Severity and exploitation remain different, context changes order, and source information still matters. The removal only reduces support for automation and implementation workflow.

\subsection{Final synthesis by research question}

The next part answers each research question. It separates direct SD-WAN evidence, transferred evidence, and standards-based requirements. It also notes agreement, disagreement, and measurement limits. A factor remains in the model only when its meaning, input, scale, age, confidence, and use can be stated clearly.

\noindent\textbf{RQ1:} The study keeps ten factors in the research model: severity, predicted exploitation, observed exploitation, exposure, reachability, privilege, blast radius, consequence, concentration, and cryptographic readiness. Vendor name and vulnerability age are metadata, not risk factors. The same applies to data age, confidence, and whether AI produced a value.

\noindent\textbf{RQ2:} Confidence is high that severity, predicted exploitation, and observed exploitation are different constructs \cite{ref4}, \cite{ref5}, \cite{ref12}, \cite{ref13}, \cite{ref15}. Evidence for external exposure is moderate and transferred; direct evidence for telemetry-freshness decay in SD-WAN is limited.

\noindent\textbf{RQ3:} Operational consequence has moderate cross-domain support, and co-exploitation evidence supports modeling dependency \cite{ref18}, \cite{ref19}. Direct SD-WAN evidence for live reachability, attainable privilege, lateral movement, blast radius, and control-plane concentration is absent; those factors remain provisional.

\noindent\textbf{RQ4:} Controller or orchestrator centrality, direct Internet access, segmentation paths, shared management services, and cryptographic trust dependencies plausibly change priority. Confidence is low because the included studies do not test these modifiers in an operational SD-WAN environment.

\noindent\textbf{RQ5:} FIPS 203, 204, and 205 define the current PQC algorithm standards \cite{ref23,ref24,ref25}. NIST transition and crypto-agility guidance supports a limited C-BOM and migration record \cite{ref26}, \cite{ref27}. Evidence is strong for inventory and transition planning. It is weak for a daily triage weight or a numeric HNDL score.

\noindent\textbf{RQ6:} The selected baseline is a transparent additive 0--100 score with normalized inputs, published weights, and a separate completeness/confidence measure. Equal weights are used only for the synthetic benchmark. Opaque learned weights, silent renormalization around missing factors, and confidence-multiplied scores are rejected for the baseline.

\subsection{Review-stage decision}

Review decision. The sources support ten distinct ideas, but not ten equally tested predictors. Severity, predicted exploitation, observed exploitation, exposure, and consequence form the first operational layer. Reachability, privilege, blast radius, and concentration need mature topology data and direct SD-WAN testing. Cryptographic readiness can be added when a reliable C-BOM exists. This paper differs from the original SLR plan. It uses a bounded review because no reproducible database search and screening log was available.

\section{PECR Framework Specification}\label{sec:framework}

PECR ranks a vulnerability on a specific asset at a specific time. It does not rank a CVE by itself. Each record identifies the vulnerability, affected asset or service, SD-WAN role, and data time. The output includes a priority score, confidence value, factor details, and any escalation flag. This prevents teams from reusing a score for a different asset or time without review.

\subsection{Scope boundary}

PECR is narrower than a general cyber-risk platform. It orders remediation for flaws in SD-WAN controllers, orchestrators, management services, branch edges, and gateways. It also covers identity services, segmentation controls, and cryptographic trust chains.

\noindent\textbf{In scope:} Technical severity, predicted and observed exploitation, exposure, reachability, privilege, blast radius, operational consequence, concentration, and cryptographic readiness.

\noindent\textbf{In scope:} Vendor-neutral inputs derived from open standards, normalized telemetry fields, topology models, public catalogs, and bounded organizational context.

\noindent\textbf{Out of scope:} Automatic patching, exploit creation, unsupported zero-day prediction, full enterprise risk measurement, and proof of cryptographic security.

\noindent\textbf{Out of scope:} Vendor product scoring, replacement of change-control approval, and claims of human explainability without an analyst study.

\subsection{Ten-factor research specification}

Table~\ref{tab:factor-specification} defines the ten factors for research use. Evidence strength decides when a factor can be adopted. It does not change the factor's 0--1 scale. Each value keeps its source, time, conversion rule, and confidence value.

\begin{table*}[!t]
\centering
\caption{Ten-Factor PECR Research Specification}
\label{tab:factor-specification}
\scriptsize
\setlength{\tabcolsep}{3pt}
\renewcommand{\arraystretch}{1.12}
\begin{tabularx}{\textwidth}{@{}p{0.032\textwidth}p{0.13\textwidth}YYY@{}}
\toprule
\textbf{No.} & \textbf{Factor} & \textbf{Normalized input} & \textbf{Evidence and freshness} & \textbf{Status} \\
\midrule
1 & Technical severity & CVSS Base score divided by 10, with version and vector retained. & NVD/vendor record; update on vector revision. & Core---direct evidence \\
2 & Predicted exploit likelihood & Versioned EPSS probability or a validated alternative on \([0,1]\). & Model version, score date, percentile, horizon. & Core---direct evidence \\
3 & Observed exploitation & Confirmed catalog or internal evidence; distinguish confirmed, absent, and unknown. & KEV, incident, honeypot, EDR/SIEM; time-stamped. & Core---direct evidence \\
4 & External exposure & Reachable service exposure, ingress path, or authenticated external access. & Flow, firewall, NAT, proxy, and asset data; decay by age. & Core---transferred evidence \\
5 & Attack-path reachability & Normalized existence, length, and control strength of a path to the affected service. & Topology and policy graph; recompute on drift. & Stage 2---provisional \\
6 & Attainable privilege & Highest credible privilege or control-plane authority gained after exploitation. & Exploit preconditions, IAM role, service account, device role. & Stage 2---provisional \\
7 & Blast radius & Normalized number and criticality of reachable segments, tenants, or dependent services. & Segmentation graph and dependency map. & Stage 2---provisional \\
8 & Operational consequence & Mission, safety, confidentiality, integrity, availability, and recovery consequence. & Business impact analysis and service tier. & Core---direct evidence \\
9 & Systemic concentration & Control-plane centrality, shared dependency, or single-point-of-failure concentration. & Architecture inventory and dependency centrality. & Stage 2---SD-WAN hypothesis \\
10 & Cryptographic readiness & Urgency from quantum-vulnerable use, confidentiality lifetime, migration lead time, and crypto-agility. & C-BOM, standards status, protocol inventory, retirement plan. & Stage 3---standards led \\
\bottomrule
\end{tabularx}
\end{table*}
\FloatBarrier

\noindent\textbf{The base calculation is transparent:} 

\begin{equation}
\mathrm{PECR}_i(t) = 100\sum_{j=1}^{10}w_jx_{ij}(t),\qquad w_j\geq 0,\qquad \sum_{j=1}^{10}w_j=1
\label{eq:pecr}
\end{equation}

\begin{equation}
C_i(t) = \sum_{j=1}^{10}w_jc_{ij}(t)
\label{eq:confidence}
\end{equation}

Equation~\eqref{eq:pecr} gives a priority score from 0 to 100. Equation~\eqref{eq:confidence} reports confidence and completeness as a separate value. PECR must not hide weak data by multiplying confidence into the score. It must also avoid silently changing weights when data are missing. The system should show the score, confidence, missing fields, and overrides together.

\subsection{Weighting, normalization, and missing data}

Weights are policy choices, not universal facts. This paper uses equal weights only as a clear and repeatable baseline. A real deployment must test weights against an external outcome. Every result should report the exact weights, tested range, and rank stability. Any threshold or nonlinear rule needs an operational reason and a factor-removal test.

\noindent\textbf{Benchmark weights:} \(w_1\) through \(w_{10}\) are each 0.10. This vector is an uncalibrated neutral baseline for the synthetic evaluation, not a recommended operational default.

\noindent\textbf{Missing data:} Preserve unknown status and calculate lower and upper score bounds by assigning the missing normalized value 0 and 1. Do not treat unknown as zero or renormalize silently. Suppress an autonomous rank when weighted completeness C is below 0.70.

\noindent\textbf{Freshness:} Keep the observed value, but reduce confidence as data become old. Use \(c=\max(0,1-\mathrm{age}/L)\). Set L to 7 days for EPSS and 24 hours for exposure or topology data. Set L to 90 days for C-BOM records. An approved local policy may use shorter periods.

\noindent\textbf{Escalation:} Confirmed exploitation creates an urgent-review flag. A control-plane asset also escalates when exposure, reachability, and consequence are each at least 0.80. Low-confidence cases are routed for evidence collection rather than silently down-ranked. The model owner approves rule changes; security operations records case-level overrides and rationale.

\subsection{Cryptographic Bill of Materials and PQC readiness}

The limited C-BOM records where public-key cryptography is used. It stores the algorithm, parameters, key and certificate lifetimes, protocol, data lifetime, owner, replacement path, and migration time. FIPS 203 defines ML-KEM for key establishment \cite{ref23}. FIPS 204 and FIPS 205 define ML-DSA and SLH-DSA for signatures \cite{ref24}, \cite{ref25}. NIST IR 8547 is still an Initial Public Draft on the manuscript date \cite{ref26}. Its transition dates must be described as proposed, not final. NIST guidance also explains how an organization can change algorithms while keeping systems secure and available \cite{ref27}.

Cryptographic readiness is not a simple yes-or-no PQC field. The score should increase when vulnerable cryptography protects long-lived confidential data. It should also increase when many branches share the same dependency or when algorithms are hard to replace. The score should decrease only when the inventory is current and a migration path has been tested. Any scoring rule must record the version and status of each standard. It must also be tested against an external outcome.

\subsection{Decision traceability and vendor neutrality}

A PECR record should show the case ID, factor values, weights, sources, dates, conversions, confidence, missing fields, overrides, score, and rank. Vendor neutrality is tested through the data schema. Two different implementations must fill the same logical fields without changing their meaning. Product APIs may provide data, but they may not define what the data mean.

\section{Evaluation Design and Worked Example}\label{sec:evaluation-design}

The evaluation has two limited goals. First, it checks whether PECR changes the CVSS order for the same synthetic cases. Second, it checks whether the calculation is traceable and stable under small weight changes. A later study must test operational value. A rank change alone is not an improvement.

\subsection{Hypotheses and success criteria}

\noindent\textbf{H1:} PECR and CVSS produce materially different rankings on matched SD-WAN instances, measured with Kendall's \(\tau_b\), \(\mathrm{MARS}\), and top-k Jaccard overlap (\(J_k\)).

\noindent\textbf{H2:} PECR improves a stated external outcome. Examples include \(\mathrm{precision}@k\) for exploitation, expert priority, avoided loss, or time to risk reduction.

\noindent\textbf{H3:} PECR produces a more complete decision trail, measured as the proportion of instances with reproducible inputs, timestamps, confidence, and factor contributions.

\noindent\textbf{H4:} The top-k queue remains acceptably stable under plausible changes in weights, telemetry age, and missing-data assumptions.

\noindent\textbf{Exploratory thresholds:} This example was not preregistered, so the thresholds are descriptive. H1 is met when \(\mathrm{MARS}\geq 1.0\) or \(J_3\leq 0.67\). H2 needs a better external outcome and cannot be tested here. H3 needs a complete factor record for every case. H4 needs the same top-three set in at least 80\% of the \(\pm 20\%\) weight combinations.

\subsection{Synthetic evaluation corpus and boundary}

Evaluation limit. The test contains five synthetic vulnerability--asset cases. They are enough to check the calculations, ranking changes, score details, and limited sensitivity. They are not production telemetry or a representative SD-WAN sample. They also do not provide external validation.

\noindent\textbf{Topology:} Five archetypes represent a controller, branch edge, orchestrator dependency, branch web interface, and management plane. No live graph or vendor implementation is used.

\noindent\textbf{Vulnerability--asset instances:} \(n=5\) synthetic cases, labeled A--E. All ten factor values are present; there are no duplicates or exclusions.

\noindent\textbf{Telemetry:} Exposure, path, privilege, blast-radius, and concentration values are scenario assignments. They are not measured IPFIX, NetFlow, event-log, routing, authentication, firewall, or proxy observations.

\noindent\textbf{External criterion:} None. The benchmark cannot test exploitation prediction, avoided loss, analyst priority, remediation outcome, or time to risk reduction.

\noindent\textbf{Data governance:} No personal, confidential, or organizational data are present. The complete synthetic matrix is published in Table~\ref{tab:synthetic-matrix}.

\noindent\textbf{Comparator:} The normalized technical-severity factor provides the CVSS-only order. Other factor values are fixed synthetic inputs and must not be interpreted as live EPSS or KEV snapshots.

\subsection{Synthetic worked example}

The five-case example shows how PECR can change a remediation queue. It uses equal weights of 0.10 only to check the calculation. All values are synthetic. A zero for observed exploitation means ``not listed in this example.'' It does not prove that exploitation has never occurred. A real system must keep unknown status and confidence.

\begin{description}
\item[\textbf{A}] Internet-reachable SD-WAN controller remote-code-execution flaw.
\item[\textbf{B}] Branch-edge command-injection flaw with confirmed exploitation.
\item[\textbf{C}] Orchestrator cryptographic dependency with long-lived confidentiality and slow migration.
\item[\textbf{D}] Branch web-interface flaw with high CVSS but limited privilege and blast radius.
\item[\textbf{E}] Management-plane authentication bypass with strong path and concentration effects.
\end{description}

\begin{table*}[!t]
\centering
\caption{Synthetic Ten-Factor Input Matrix}
\label{tab:synthetic-matrix}
\scriptsize
\setlength{\tabcolsep}{3pt}
\renewcommand{\arraystretch}{1.12}
\begin{tabularx}{\textwidth}{@{}Ycccccc@{}}
\toprule
\textbf{Factor} & \textbf{Weight} & \textbf{A} & \textbf{B} & \textbf{C} & \textbf{D} & \textbf{E} \\
\midrule
Technical severity (TS) & 0.10 & 0.98 & 0.88 & 0.75 & 0.90 & 0.81 \\
Predicted likelihood (PL) & 0.10 & 0.92 & 0.65 & 0.05 & 0.08 & 0.35 \\
Observed exploitation (OE) & 0.10 & 1.00 & 1.00 & 0.00 & 0.00 & 0.00 \\
External exposure (EX) & 0.10 & 0.90 & 1.00 & 0.40 & 0.70 & 0.60 \\
Attack-path reachability (AR) & 0.10 & 0.95 & 0.80 & 0.65 & 0.25 & 0.90 \\
Attainable privilege (AP) & 0.10 & 0.90 & 0.75 & 0.60 & 0.20 & 1.00 \\
Blast radius (BR) & 0.10 & 0.95 & 0.65 & 0.85 & 0.10 & 0.90 \\
Operational consequence (OC) & 0.10 & 1.00 & 0.80 & 0.95 & 0.30 & 0.90 \\
Systemic concentration (SC) & 0.10 & 1.00 & 0.70 & 1.00 & 0.10 & 0.90 \\
Cryptographic readiness (CR) & 0.10 & 0.20 & 0.40 & 0.95 & 0.00 & 0.10 \\
PECR score from Equation~\eqref{eq:pecr} & --- & 88.0 & 76.3 & 62.0 & 26.3 & 64.6 \\
\bottomrule
\end{tabularx}
\end{table*}

CVSS orders the instances A, D, B, E, C. Equal-weight PECR orders them A, B, E, C, D. Table~\ref{tab:rank-comparison} shows the resulting rank movement.

\begin{table*}[!t]
\centering
\caption{Synthetic CVSS--PECR Rank Comparison}
\label{tab:rank-comparison}
\footnotesize
\setlength{\tabcolsep}{3pt}
\renewcommand{\arraystretch}{1.12}
\begin{tabularx}{\textwidth}{@{}cp{0.42\textwidth}ccc@{}}
\toprule
\textbf{ID} & \textbf{Scenario} & \textbf{CVSS rank} & \textbf{PECR rank} & \textbf{|Shift|} \\
\midrule
A & Controller RCE & 1 & 1 & 0 \\
B & Exploited branch-edge injection & 3 & 2 & 1 \\
C & Orchestrator crypto dependency & 5 & 4 & 1 \\
D & Limited branch web flaw & 2 & 5 & 3 \\
E & Management-plane auth bypass & 4 & 3 & 1 \\
\bottomrule
\end{tabularx}
\end{table*}
\FloatBarrier

For this example, \(\mathrm{MARS}=(0+1+1+3+1)/5=1.2\) positions. The top-three lists share A and B, so \(J_3=2/4=0.50\). Seven pairs agree and three disagree. With no ties, Kendall's \(\tau_b=(7-3)/10=0.40\). There are \(5!=120\) possible orders. The exact two-sided p-value for \(\lvert\tau\rvert\geq 0.40\) is \(58/120=0.483\). The result meets the descriptive threshold. However, five cases cannot support a significant result or a claim of operational value.

\subsection{Reproducible evaluation procedure}

\begin{enumerate}
\item Freeze the vulnerability, asset, topology, telemetry, CVSS, EPSS, KEV, C-BOM, and external-outcome snapshots; record versions and hashes.
\item Apply the final normalization and missing-data rules without viewing the outcome labels.
\item Compute CVSS and PECR ranks on the same instances and retain decomposed score records.
\item Report Kendall's \(\tau_b\), \(\mathrm{MARS}\), top-k Jaccard (\(J_k\)), confidence intervals, tie handling, and rank-change explanations.
\item Test external outcome performance with prespecified metrics and compare against CVSS and secondary baselines.
\item Run ablation, weight, freshness, and missing-data sensitivity analyses; report every setting.
\item If the paper claims human explainability, run a separate approved analyst study. Measure understanding, trust, decision time, and decision quality.
\end{enumerate}

\section{Synthetic Evaluation Results}\label{sec:evaluation-results}

Result limit. These results apply only to the synthetic matrix and equal-weight model. They show that PECR can produce a different and traceable order under stated assumptions. They do not show better real-world remediation.

\subsection{Dataset and execution summary}

The test has five synthetic cases and five SD-WAN role types. All 50 factor values are present, so missingness is 0\%. There is no collection period, branch count, production site, or excluded case. Tables~\ref{tab:synthetic-matrix} and~\ref{tab:rank-comparison}, Equations~\eqref{eq:mars}--\eqref{eq:confidence}, and the weight rule define the full calculation.

\subsection{Ranking divergence}

CVSS ranked the cases A--D--B--E--C. PECR ranked them A--B--E--C--D. Kendall's \(\tau_b\) was 0.40 with no ties, and the exact two-sided p-value was 0.483. Rank shifts were 0, 1, 1, 3, and 1. \(\mathrm{MARS}\) was 1.2, the median shift was 1, and the largest shift was 3. \(J_3\) was 0.50. Case D fell from second to fifth. Its high severity was balanced by low reachability, privilege, blast radius, consequence, and concentration.

\subsection{External outcome and operational utility}

No external outcome or secondary operational baseline exists for the synthetic corpus. H2 is therefore not tested. The results cannot establish improved precision, avoided loss, lower workload, faster remediation, or fewer false negatives. The only supported claim is rank divergence under transparent assumptions.

\subsection{Traceability, ablation, and sensitivity}

All five synthetic records are complete and show every score component. The base order stayed the same in eight of ten single-factor removal tests. Removing predicted likelihood or privilege changed the middle order to A--B--C--E--D. Doubling one factor kept the base order in nine of ten tests. Doubling cryptographic readiness moved C above E. The exact order and top-three set stayed the same in 87.5\% of the 1,024 weight combinations. In the other cases, \(J_3\) fell to 0.50. We did not test old or missing telemetry because the matrix is static and complete.

\begin{table*}[!t]
\centering
\caption{Synthetic Evaluation Results}
\label{tab:evaluation-results}
\scriptsize
\setlength{\tabcolsep}{3pt}
\renewcommand{\arraystretch}{1.12}
\begin{tabularx}{\textwidth}{@{}p{0.16\textwidth}p{0.23\textwidth}Y@{}}
\toprule
\textbf{Outcome} & \textbf{Statistic or test} & \textbf{Result} \\
\midrule
Rank association & Kendall's \(\tau_b\); exact test; tie handling & \(\tau_b=0.40\); exact two-sided \(p=0.483\); no ties \\
Rank movement & \(\mathrm{MARS}\); median and maximum shift & \(\mathrm{MARS}=1.2\); median = 1; maximum = 3 \\
Top-k overlap & Jaccard at the reported k & \(J_3=0.50\); shared cases: A and B \\
External validity & External outcome metric and effect size & Not tested; the synthetic corpus has no ground truth \\
Traceability & Complete records and recomputation & 5/5 complete by construction; no independent user study \\
Robustness & Ablation and bounded weight perturbation & Base order retained in 8/10 ablations and 87.5\% of 1,024 \(\pm 20\%\) weight corners \\
Operational impact & Workload, SLA, loss, or time-to-risk metric & Not measured; no production remediation process was observed \\
\bottomrule
\end{tabularx}
\end{table*}

\subsection{Final answer to the evaluation question}

Answer to the ERQ. Yes. Equal-weight PECR produced a different ranking from CVSS in the five cases. The results were \(\tau_b=0.40\), \(\mathrm{MARS}=1.2\), and \(J_3=0.50\). The score is computationally traceable because every factor, weight, and assumption is visible. The test does not show better operational results or better human understanding. The finding applies only to the complete synthetic matrix and the tested weight range.

\section{Adoption and Governance Model}\label{sec:adoption}

PECR should be used as a managed decision process, not as a one-time score. An organization can start with asset inventory and clear decision records. It can then add live exposure, topology, cryptographic dependencies, and outcome-based calibration. Table~\ref{tab:adoption-model} presents these steps without requiring full telemetry on the first day.

\begin{table*}[!t]
\centering
\caption{PECR Adoption Maturity Model}
\label{tab:adoption-model}
\scriptsize
\setlength{\tabcolsep}{3pt}
\renewcommand{\arraystretch}{1.12}
\begin{tabularx}{\textwidth}{@{}p{0.13\textwidth}p{0.35\textwidth}Y@{}}
\toprule
\textbf{Stage} & \textbf{Capability} & \textbf{Exit criterion} \\
\midrule
1. Inventory & Map vulnerabilities to assets, SD-WAN roles, owners, and CVSS/EPSS/KEV snapshots. & At least 95\% of in-scope assets have current ownership and vulnerability-asset mapping. \\
2. Exposure & Add normalized external exposure and telemetry freshness with source timestamps. & Exposure evidence and age are available for the operational queue; unknowns are explicit. \\
3. Path and consequence & Add topology, reachability, privilege, blast radius, consequence, and concentration. & Rank records reproduce from a versioned graph and approved business-impact inputs. \\
4. Cryptographic readiness & Add the bounded C-BOM, data lifetime, migration lead time, and crypto-agility evidence. & Critical trust chains have owners, algorithm inventory, and tested transition plans. \\
5. Validated operation & Calibrate against outcomes; monitor drift, stability, overrides, workload, and decision quality. & The model meets preregistered outcome and robustness thresholds across more than one implementation. \\
\bottomrule
\end{tabularx}
\end{table*}

\subsection{Governance and operational controls}

\noindent\textbf{Model owner:} Approves factor definitions, weights, thresholds, missing-data rules, and version changes.

\noindent\textbf{Data owners:} Attest to asset, telemetry, topology, business-impact, and C-BOM coverage and freshness.

\noindent\textbf{Security operations:} Reviews the queue, records overrides and rationale, and reports remediation outcomes.

\noindent\textbf{Change advisory authority:} Decides remediation timing when operational risk conflicts with vulnerability urgency.

\noindent\textbf{Independent assurance:} Reproduces sample decisions, audits evidence provenance, and reviews drift and bias.

Adoption measures should cover inventory, missing factors, data age, rank stability, and override rate. They should also cover decision time, SLA results, urgent-queue size, missed priorities, and analyst acceptance. Each evaluation must use one fixed model version. Important model changes require a new validation.

\subsection{Vendor-neutral implementation test}

A system is vendor-neutral only when factor meanings stay the same across data adapters. Testing should use the same schema with at least two SD-WAN or similar network products. The test should compare missing data and conversion rules. It should also list any field that depends on a proprietary feature. A vendor incident may inform one case, but it cannot define a universal factor.

\section{Discussion}\label{sec:discussion}

\subsection{Answer to the evaluation question}

The review shows that severity, exploit likelihood, and organizational context answer different questions \cite{ref4,ref5,ref6,ref7,ref8,ref9,ref10,ref11,ref12,ref13,ref14,ref15}. The synthetic test shows that PECR can reorder the same cases. Case D fell three places because its high severity did not come with strong path, privilege, blast-radius, consequence, or concentration signals. Cases B and E moved up. Their exploitation, reachability, privilege, and management-plane context added information that CVSS did not include.

PECR can support traceable decisions by showing factors, dates, sources, confidence, and score contributions. We can test this through record completeness and repeatable calculations. We should not call PECR easier for people to understand without an analyst study. That study should measure understanding, trust, and decision quality.

\subsection{Coverage of the research questions}

\begin{table*}[!t]
\centering
\caption{Answers and Evidence Boundaries for the Research Questions}
\label{tab:rq-answers}
\scriptsize
\setlength{\tabcolsep}{3pt}
\renewcommand{\arraystretch}{1.12}
\begin{tabularx}{\textwidth}{@{}p{0.10\textwidth}YY@{}}
\toprule
\textbf{Question} & \textbf{Bounded evidence and synthetic result} & \textbf{Conclusion and residual boundary} \\
\midrule
PRQ / RQ1 & The corpus supports distinct constructs for severity, exploitation, exposure, consequence, topology, and cryptographic readiness \cite{ref4,ref5,ref6,ref7,ref8,ref9,ref10,ref11,ref12,ref13,ref14,ref15,ref16,ref17,ref18,ref19,ref20,ref21,ref22,ref23,ref24,ref25,ref26,ref27}. & Retain ten factors as a staged research specification. Evidence strength is unequal, and the set is not claimed to be optimal. \\
RQ2 & Severity, predicted exploitation, and observed exploitation are distinct signals \cite{ref4}, \cite{ref5}, \cite{ref12}, \cite{ref13}, \cite{ref15}. & High confidence for separating these signals; moderate for exposure; limited direct evidence for telemetry-freshness rules. \\
RQ3 & Consequence and co-exploitation have adjacent support \cite{ref18}, \cite{ref19}; contextual ordering is supported \cite{ref6,ref7,ref8,ref9,ref10}. & Direct SD-WAN reachability, privilege, blast-radius, and concentration measurements remain unvalidated. \\
RQ4 & Software-defined, hybrid-cloud, IoT, and automotive studies provide transferable architecture concepts \cite{ref6,ref7,ref8}, \cite{ref16}, \cite{ref22}. & Controller centrality, direct Internet access, and segmentation are plausible modifiers, but the corpus contains no direct SD-WAN validation. \\
RQ5 & FIPS 203--205 define current PQC standards; NIST guidance supports transition inventory and crypto agility \cite{ref23,ref24,ref25,ref26,ref27}. & Standards justify a bounded C-BOM and migration-readiness construct, not a validated daily triage weight. \\
RQ6 & Prior work supports contextual weights, sensitivity analysis, and explicit evidence provenance \cite{ref6}, \cite{ref8,ref9,ref10}, \cite{ref14}. & Use a transparent additive score with a separate confidence measure; equal weights remain an illustrative baseline. \\
ERQ & The order changed from A--D--B--E--C to A--B--E--C--D. The results were \(\tau_b=0.40\), \(\mathrm{MARS}=1.2\), and \(J_3=0.50\). & PECR can produce a measurably different and decomposable ranking. Better operational outcomes and human explainability were not tested. \\
\bottomrule
\end{tabularx}
\end{table*}
\FloatBarrier

\subsection{Evidence boundary and unresolved gaps}

This review reports published results and does not independently repeat the original experiments. Claims from abstracts are limited to the information in those abstracts. Reviews are kept separate from primary studies to avoid double counting. The DOI-based study set is not exhaustive. One reviewer completed the quality check, and the equal-weight test is synthetic. PRISMA is a reporting guide \cite{ref1}, but this project does not have the retrieval and screening counts needed for a PRISMA claim. We therefore call it a bounded multivocal review based on software-engineering guidance \cite{ref2}, \cite{ref3}.

\subsection{Interpretation and practical trade-offs}

More factors do not automatically make PECR better. Its main value is that readers can see why a case changed position. This visibility also shows the trade-offs. Large exploitation or exposure weights may overreact to noisy data. Small weights may miss an exploited flaw with moderate severity. Cryptographic readiness may raise a long-term confidentiality risk. However, it may push down an urgent and exploitable flaw when both use one queue. A two-lane queue or an escalation rule may work better than one unrestricted score. Direct use in SD-WAN is still uncertain. This study did not test live topology, vendor adapters, analyst understanding, or remediation results. SSVC decision trees, two-stage gates, and learning-to-rank models are reasonable alternatives. PECR should be used operationally only after testing these alternatives on the same data.

\section{Methodological Risks and Validity Threats}\label{sec:validity}

\noindent\textbf{Corpus coverage:} The DOI-led, bounded evidence set may omit relevant databases, non-English work, unpublished negative results, and product-independent SD-WAN studies.

\noindent\textbf{Evidence asymmetry:} PQC readiness depends mainly on standards and transition guidance, while its daily vulnerability-prioritization value remains empirically untested.

\noindent\textbf{Reviewer bias:} One reviewer selected, extracted, and appraised the bounded corpus. No kappa statistic or adjudication process is available.

\noindent\textbf{Construct validity:} Decision traceability, analyst comprehension, remediation utility, and risk reduction are distinct outcomes and must not be collapsed.

\noindent\textbf{Temporal validity:} Standards, catalogs, and predictive models can change. Their version and status must be rechecked at submission and in every operational run.

\noindent\textbf{Synthetic-data validity:} Five designed cases can demonstrate mechanics and failure modes but cannot estimate population effects, calibration, workload, or field performance.

\noindent\textbf{Unit-of-analysis mismatch:} Vendor, vulnerability, asset, and attack-path rankings are not directly comparable without an explicit mapping.

\noindent\textbf{Source-depth limitation:} An abstract can support only abstract-level claims; it cannot establish unreported baselines, sampling decisions, statistical tests, or operational effectiveness.

\noindent\textbf{Weight and model drift:} Equal weights are not calibrated. Operational evaluations must freeze factor definitions, EPSS version, transformations, weights, and retrieval dates.

\noindent\textbf{Implementation validity:} Vendor neutrality is specified at the schema boundary but was not tested with two independent SD-WAN implementations.

\section{Conclusion}\label{sec:conclusion}

The review supports a vendor-neutral PECR research model. It keeps severity, exploitation, exposure, topology, privilege, blast radius, consequence, concentration, and cryptographic readiness as separate factors. Evidence is strongest for separating severity from exploitation and adding organizational context. Evidence is weaker for live SD-WAN attack paths and a daily PQC migration weight. A clear additive score with a separate confidence value is a reasonable starting point. However, its weights are policy choices, not universal constants.

In the five-case test, CVSS and PECR produced different orders. The results were \(\tau_b=0.40\), exact \(p=0.483\), \(\mathrm{MARS}=1.2\), and \(J_3=0.50\). The PECR order stayed the same in 87.5\% of the tested weight combinations. This answers the narrow question: PECR can produce a different and traceable ranking. It does not show that the ranking is more accurate, safer, easier to understand, or better in practice. The main contribution is a clear model, a worked calculation, and a path for future testing with real SD-WAN data.

\FloatBarrier
\section*{Declarations}

\noindent\textbf{Funding Statement:} This research received no external funding.

\noindent\textbf{Conflict of Interest Statement:} The author declares no conflicts of interest.

\noindent\textbf{CRediT Author Statement:} Saeed Alam: Conceptualization, methodology, investigation, formal analysis, writing---original draft, writing---review and editing, and project administration.

\noindent\textbf{Data Availability Statement:} All literature records are publicly identified in the reference list. The complete synthetic evaluation matrix and ranks are reported in Tables~\ref{tab:synthetic-matrix} and~\ref{tab:rank-comparison}. No production network or personal data were used.

\noindent\textbf{Code Availability Statement:} The calculations are fully specified by Equations~\eqref{eq:mars}--\eqref{eq:confidence}, Tables~\ref{tab:synthetic-matrix}--\ref{tab:evaluation-results}, and the stated perturbation procedure.

\noindent\textbf{Ethics Approval:} Not applicable. The work used published literature and synthetic data only. Any future operational study requires separate institutional review.

\noindent\textbf{AI-Use Disclosure:} The author utilized AI-assisted editing tools for language polishing, LaTeX formatting, document structuring, and calculation auditing. The author independently verified all outputs and maintains full responsibility for the contents of the manuscript.

\end{document}